\documentclass[sigconf, nonacm]{acmart}
\usepackage{graphicx, dblfloatfix}
\usepackage{balance}
\usepackage{graphicx}
\usepackage{textcomp}
\usepackage{booktabs}
\usepackage{xcolor}
\usepackage{tabularx}
\usepackage{footnote}
\usepackage{array}
\usepackage[a-1b]{pdfx}
\usepackage{multirow}
\usepackage{url}
\usepackage{capt-of}
\usepackage{siunitx}
\usepackage{dsfont}
\usepackage{subcaption}
\usepackage{bm}
\usepackage[inline]{enumitem}
\newcolumntype{P}[1]{>{\centering\arraybackslash}p{#1}}

\newcommand\vldbdoi{10.14778/3476249.3476280}
\newcommand\vldbpages{2283 - 2295}
\newcommand\vldbvolume{14}
\newcommand\vldbissue{11}
\newcommand\vldbyear{2021}
\newcommand\vldbauthors{\authors}
\newcommand\vldbtitle{\shorttitle} 
\newcommand\vldbavailabilityurl{}
\newcommand\vldbpagestyle{empty} 

\newcommand{\para}[1]{\smallskip\noindent\textit{#1.}}
\newcommand{\eat}[1]{}

\begin{document}

\title{Real-World Trajectory Sharing with Local Differential Privacy}

\author{Teddy Cunningham}
\affiliation{
  \institution{University of Warwick}
  \city{Coventry}
  \country{United Kingdom}
  \postcode{CV4 7AL}
}
\email{teddy.cunningham@warwick.ac.uk}

\author{Graham Cormode}
\affiliation{
  \institution{University of Warwick}
  \city{Coventry}
  \country{United Kingdom}
  \postcode{CV4 7AL}
}
\email{g.cormode@warwick.ac.uk}

\author{Hakan Ferhatosmanoglu}
\affiliation{
  \institution{University of Warwick}
  \city{Coventry}
  \country{United Kingdom}
  \postcode{CV4 7AL}
}
\email{hakan.f@warwick.ac.uk}

\author{Divesh Srivastava}
\affiliation{
  \institution{AT\&T Chief Data Office}
  \city{Bedminster, NJ}
  \country{USA}
}
\email{divesh@att.com}

\renewcommand{\shortauthors}{Cunningham, et al.}

\begin{abstract}
Sharing trajectories is beneficial for many real-world applications, such as managing disease spread through contact tracing and tailoring public services to a population's travel patterns. 
However, public concern over privacy and data protection has limited the extent to which this data is shared.  
Local differential privacy  enables data sharing in which users share a perturbed version of their data, but existing mechanisms fail to incorporate user-independent public knowledge (e.g., business locations and opening times, public transport schedules, geo-located tweets).  
This limitation makes mechanisms too restrictive, gives unrealistic outputs, and ultimately leads to low practical utility.
To address these concerns, we propose a local differentially private mechanism that is based on perturbing hierarchically-structured, overlapping $n$-grams (i.e., contiguous subsequences of length $n$) of trajectory data.  
Our mechanism uses a multi-dimensional hierarchy over publicly available external knowledge of real-world places of interest to improve the realism and utility of the perturbed, shared trajectories.
Importantly, including real-world public data does not negatively affect privacy or efficiency.  
Our experiments, using real-world data and a range of queries, each with real-world application analogues, demonstrate the superiority of our approach over a range of alternative methods.
\end{abstract}

\maketitle

\pagestyle{\vldbpagestyle}
\begingroup\small\noindent\raggedright\textbf{PVLDB Reference Format:}\\
\vldbauthors. \vldbtitle. PVLDB, \vldbvolume(\vldbissue): \vldbpages, \vldbyear.\\
\href{https://doi.org/\vldbdoi}{doi:\vldbdoi}
\endgroup
\begingroup
\renewcommand\thefootnote{}\footnote{\noindent
This work is licensed under the Creative Commons BY-NC-ND 4.0 International License. Visit \url{https://creativecommons.org/licenses/by-nc-nd/4.0/} to view a copy of this license. For any use beyond those covered by this license, obtain permission by emailing \href{mailto:info@vldb.org}{info@vldb.org}. Copyright is held by the owner/author(s). Publication rights licensed to the VLDB Endowment. \\
\raggedright Proceedings of the VLDB Endowment, Vol. \vldbvolume, No. \vldbissue\ %
ISSN 2150-8097. \\
\href{https://doi.org/\vldbdoi}{doi:\vldbdoi} \\
}\addtocounter{footnote}{-1}\endgroup

\ifdefempty{\vldbavailabilityurl}{}{
\vspace{.3cm}
\begingroup\small\noindent\raggedright\textbf{PVLDB Artifact Availability:}\\
The source code, data, and/or other artifacts have been made available at \url{\vldbavailabilityurl}.
\endgroup
}

\section{Introduction}
\label{s:intro}
Sharing trajectories has obvious benefits for many real-world applications, including managing disease spread through contact tracing, and tailoring public services (such as bus routes) to a population's travel patterns.  
However, widespread public concern over privacy and data protection has limited the extent to which this data is currently shared.  
Differential privacy (DP) is an increasingly popular technique for publishing sensitive data with provable guarantees on individual privacy. 
In contrast to centralized DP, local differential privacy (LDP) allows users to share a perturbed version of their data, thus allaying fears of an untrusted data collector.

Although LDP provides a more practical setting and more attractive privacy properties, its mechanisms often have lower utility due to its stronger privacy requirements.
This is (in part) because existing mechanisms fail to incorporate the wide range of real-world knowledge that is publicly available.
This is a major shortcoming, especially as a wealth of accessible, open source information about the real world exists: detailed mapping data describing roads and points of interest; transit schedules; business opening hours; and unstructured user-generated data, in the form of reviews, check-ins, photos, and videos.
These provide a rich and detailed (if somewhat non-uniform) description of the real world within which people navigate their lives. 
Within this context, approaches to data sharing that rely on crude abstractions, such as describing counts within uniform grids, appear simplistic. 
Moreover, they lead to synthetic data that fails to respect common sense: showing movement patterns that cross a mountain range as if there was a highway through it, or trajectories in which travelers visit a sports stadium in the middle of the night.  
We argue that, to be of value, efforts to share trajectories must more explicitly model the real world, and combine the private data with public information. 
To better capture realistic behavior, we propose solutions that include a wide range of external knowledge in a utility-enhancing manner, and show empirically that including real-world information greatly improves the utility of the perturbed data.

External knowledge can be incorporated in two ways.  
The first is a series of (deterministic) constraints that simply state whether one instance is feasible or not (e.g., someone cannot be `located' in the sea).
The second (probabilistic) approach is to make certain outputs more likely than others.  
Indeed, the second approach relates to another limitation of existing LDP mechanisms, wherein many assume equal sensitivity across data points.  
That is, with respect to place A, they treat places B and C to be equally sensitive, even if place A is much `closer' to place B than to place C.  
Although a number of existing approaches instead use non-uniform probabilities in a utility-minded manner \cite{Andres2013, Alvim2018, Acharya2019, Gu2020}, these works use relaxed definitions of $\epsilon$-LDP, which is unsatisfying.

Our framework includes both of these approaches. 
First, a 'reachability' constraint ensures adjacent points in a trajectory can be reached in the respective time gap.  
Second, we use a multi-attributed distance function that ensures semantically similar locations are more likely to be returned by a perturbation mechanism.
However, privately perturbing trajectories in the local setting while incorporating real-world knowledge effectively is non-trivial.  
Furthermore, we go further than existing methods by proposing a solution that satisfies the strict requirements of $\epsilon$-LDP.  

Our first mechanism -- which models trajectories as individual points in high-dimensional space -- can be seen as the elegant, `global' solution. 
However, its time and space complexity makes it computationally infeasible in most scenarios, which leads us to introduce our more efficient and scalable solution, based on perturbing overlapping, hierarchically-structured $n$-grams (i.e., contiguous subsequences of length $n$) of trajectory data.  
$n$-gram perturbation allows us to capture the spatio-temporal relationship between adjacent points, while remaining computationally feasible.
Moreover, using \textit{overlapping} $n$-grams allows us to capture more information for each point, whilst continuing to satisfy LDP.
Our semantic distance function incorporates a rich set of public knowledge to adjust the probability of certain perturbations in a utility-enhancing manner. 
We also exploit the (publicly-known) hierarchies that are inherent in space, time, and category classifications to structure $n$-grams in a multi-dimensional hierarchy, which has notable benefits for utility. 
Exploiting a hierarchically-structured space in this manner also reduces the scale of the problem, which ensures that our solution is scalable for large urban datasets.
Finally, we use optimization techniques to reconstruct the optimal realistic output trajectory from the set of perturbed $n$-grams.

We compare our mechanism to a number of alternative approaches by comparing the real and perturbed trajectory sets, and answering a range of application-inspired queries.  
Our mechanism produces perturbed trajectory sets that have high utility, preserve each location's category better than alternatives in all settings, and also preserve the temporal location of hotspots at a range of granularities.
Our solution scales well with city size (unlike some other baselines), while remaining efficient and accurate.

In summary, the main contributions of our work are:
\begin{itemize}
    \item an outline of the global solution, which we argue to be computationally infeasible in most cases;
    \item a robust, scalable, and efficient mechanism for perturbing spatio-temporal trajectory data in a way that satisfies $\epsilon$-LDP;
    \item a method for integrating public knowledge into private mechanisms to give significant utility improvements with no cost to privacy; and
    \item extensive empirical analysis, including through a range of queries, that indicate our work's relevance in addressing important data analytics problems in a private manner.
\end{itemize}

We consider related work (Section~\ref{s:related-work}) before formally introducing our problem and the guiding principles for our solution (Section~\ref{s:problem-motivation}). 
We introduce key definitions that are integral to our solution in Section \ref{s:definitions}.  
We begin Section \ref{s:solution} with the `global' solution before outlining our hierarchical, $n$-gram-based solution.  
We also outline the semantic distance function and discuss the efficiency of our mechanism. 
We set out our experiments in Section \ref{s:expts} and present the results in Section \ref{s:results}.  
We discuss future work in Section \ref{s:conc}.
\section{Related Work}
\label{s:related-work}

Differential privacy \cite{Dwork2006} has become the de facto privacy standard.
While centralized DP assumes data aggregators can be trusted, LDP \cite{Duchi2013} assumes that aggregators cannot be trusted and relies on data providers to perturb their own data.  
Many early LDP mechanisms \cite[e.g.,][]{Erlingsson2014, Wang2017, Bassily2017} assume all data points have equal sensitivity (i.e., the probability of any other data point being returned is equal), which can be unrealistic in  practical settings, especially for spatial data.

There have been a number of recent relaxations of (L)DP to allow perturbation probabilities to be non-uniform across the domain.  
In $d_\chi$-privacy \cite{Chatzikokolakis2013}, and its location-specific variant geoindistinguishability \cite{Andres2013}, the indistinguishability level between any two inputs is a function of the distance between them.  
This concept has since been generalized to any metric, and extended to the local setting to give metric-LDP \cite{Alvim2018}.  
Context-aware LDP \cite{Acharya2019} goes further by allowing an arbitrary (non-metric) measure of similarity between points, and input-discriminative LDP \cite{Gu2020} assigns each data point its own privacy level.  
Other relaxations to LDP rely on the provision of some additional information.  
For example, personalized LDP \cite{Chen2016} lets users specify a desired privacy level, whereas local information privacy \cite{Jiang2018} utilizes knowledge of users' priors.  

Location data privacy (surveyed in~\cite{Jiang2021}) has received a reasonable degree of attention in both centralized and local DP studies, (summarized in \cite{Errounda2020}).
In addition to the aforementioned LDP relaxations, there have been some specific relaxations to LDP in the location domain (summarized in~\cite{Machanavajjhala2018}).
Applying (L)DP techniques to trajectory data, however, is less well-studied.
Most early work used trajectory data to answer common queries
\cite[e.g.,][]{Chen2012, Bonomi2013, Li2017a}, which focus on returning summary level statistics, as opposed to individual-level data, which gives end users more flexibility.
More recent DP-related work has focused on publishing and synthesizing differentially private trajectories \cite[e.g.,][]{He2015, Gursoy2019a, Gursoy2020}.

There has been DP-related work that considers user-specific context in which user priors are utilized \cite[e.g.,][]{Kifer2012, Li2013, Jiang2018}, and \citet{Desfontaines2020} study the notion of DP with `partial knowledge' for a user.
Finally, \citet{Cunningham2021} use publicly available geographic knowledge to improve the quality of private synthetic location data (in the centralized setting). 
However, we are the first to explicitly use external domain knowledge (i.e., user-independent prior information that is known to all, such as the locations, business opening hours, etc.) in the local setting to enhance utility. 

In summary, our work introduces a rigorous and unified LDP-based and utility-minded mechanism for publishing individual-level trajectory data that incorporates a wide range of public knowledge.  

\section{Problem Motivation}
\label{s:problem-motivation}
Imagine a city in which each resident visits a number of places of interest (POIs) each day.  
These POIs link together in a time-ordered sequence to form a trajectory.  
The city's government wishes to learn aggregate information on where residents are traveling but, wary of governmental oversight, many residents are unwilling to share their entire trajectories truthfully. However, they are willing to share a slightly perturbed version of their trajectory, especially if it came with privacy guarantees.  
Hence, we wish to create a mechanism for users to share their trajectories in a privacy-preserving way, whilst ensuring that the shared trajectories preserve the major underlying patterns and trends of the real data at the aggregate level. 
We now describe the three principles that motivate and guide our solution: protecting privacy, enhancing utility, and ensuring efficiency.  

\para{Protecting Privacy}
\label{ss:protecting-privacy}
Our primary aim is to protect the individual privacy of each user so that she has plausible deniability within the dataset.
We seek to achieve this by perturbing each user's trajectory in order to satisfy the requirements of DP.  Specifically, we will use LDP wherein the data aggregator is not trusted.
We assume that each user shares one trajectory each, and it is shared at the end of the data collection period.
We discuss the privacy implications of these assumptions in Section \ref{ss:privacy-analysis}.

\para{Enhancing Utility by Incorporating External Knowledge}
\label{ss:enhancing-utility}
Although the primary aim of the mechanism is to preserve privacy, our practical goal is to ensure that the perturbed trajectories have high utility.
Information to preserve to ensure high utility can range from hotspot information to co-location patterns and previous travel history. 
We argue that an important way to boost utility is to link the probability of perturbation from one location to any other with the semantic distance between the two locations.  
That is, one is more likely to be perturbed to another location if it is more semantically similar to its current location.  

Furthermore, traditional (L)DP models impose strong privacy guarantees to protect against external information being used in an adversarial attack.  
However, in real-world applications of (L)DP, we argue that these protections can be too strong and can negatively affect the utility of the output dataset.  
To improve utility, we propose incorporating a range of \textit{publicly known} external information to influence the output of our mechanisms.  
Examples of this type of information include business opening hours, sports teams schedules, and published maps.  
This knowledge can be used to influence the likelihood of certain perturbations, with the aim of boosting realism and utility.
As this knowledge is publicly available, it is assumed adversaries have access to it.

\para{Ensuring Efficiency}
\label{ss:ensuring-efficiency}
As our solutions utilize a wide range of public information from the real world, it is equally important that our solution can be applied to real world settings, at scale.  
Consequently, we complement our privacy and utility goals with the desire for our solution to be efficient and scalable for large urban datasets.

\para{Applications}
\label{ss:applications}
Our work focuses on perturbing trajectories such that aggregate statistics are preserved as much as possible, which leads to many important end applications of our work.
A notable (and timely) one is the idea of \textit{societal} contact tracing that seeks to identify the places and times in which large groups of people meet (so-called `superspreading' events), as opposed to chance encounters between individuals.  
Knowledge of such events can be used for location-specific announcements and policy decisions. 
Other applications include advertising and provision of public services.  
For example, if a city council can identify popular trip chains among residents, they can improve the public transport infrastructure that links these popular places.
Likewise, if a restaurant owner knows that many museum-goers eat lunch out after visiting a museum, she may consider advertising near museums.

\section{Definitions}
\label{s:definitions}
In this section, we introduce necessary notation and definitions; commonly-used notation is summarized in Table \ref{tab:notation}.
We denote a set of POIs, as $\mathcal{P}$, where an individual POI is denoted by $p_i$.  
Each POI $p_i \in \mathcal{P}$ has a number of attributes -- $\alpha_{it}, \beta_{it}$, etc. -- associated with it, which could represent the popularity, privacy level, category, etc. of the POI, and they can vary temporally.
We quantize the time domain into a series of timesteps $t$, the size of which is controlled by the time granularity, $g_t$.  
For example, if $g_t = 5$ minutes, the time domain would be: $T = \{$...10:00, 10:05, 10:10,...$\}$.  

We define a trajectory, $\tau$, at the POI level as a sequence of POI-timestep pairs such that $\tau = \{(p_1, t_1), ... (p_i, t_i), ..., (p_{|\tau|}, t_{|\tau|})\}$, where $|\tau|$ denotes the number of POI-timestep pairs in a trajectory (i.e., its length). 
For each trajectory, we mandate that $t_{i+1} > t_i$ (i.e., one cannot go back in time, or be in two places at once). 
Each trajectory is part of a trajectory set, $\mathcal{T}$.
Perturbed trajectories and trajectory sets are denoted as $\hat{\tau}$ and $\widehat{\mathcal{T}}$, respectively.  

We use combined space-time-category (STC) hierarchical partitions in which we assign POIs to different STC regions.
$r_\chi$ denotes an individual region where $\chi$ denotes the dimension of the region (i.e., $s$ for space, $t$ for time, etc.). 
Regions can be combined to form STC regions $r_{stc}$ and $\mathcal{R}_\chi$ denotes region sets.  
A trajectory can be represented on the region level as $\tau = \{r_1, ... r_i, ..., r_{|\tau|}\}$ whereby $r_i$ represents the $i$-th STC region in the trajectory.
Consider the first point in a trajectory being \{Central Park, 10:54am\}.
This might give $r_s =$ \{Upper Manhattan\}, $r_c =$ \{Park\}, and $r_t =$ \{10-11am\}, leading to $r_{stc} = r_1 = $ \{Upper Manhattan, 10-11am, Park\}.

Chaining regions (or POIs) together forms $n$-grams, denoted as $w^n$, whereby: $w^n = \{r_i, \ldots, r_{i+n-1}\}$. 
$\mathcal{W}^n$ is the set of all possible $n$-grams.
We use $\tau(a,b)$ to specify a sub-sequence of $\tau$ such that: $\tau(a,b) = \{r_a,..., r_i,... r_b\}$, where $a$ and $b$ are the indices of $\tau$.  
For example, $\tau(1,3)$ denotes the first three STC regions (or POI-timestep pairs) of $\tau$. 
$d_s(p_i, p_j)$ denotes the physical distance between $p_i$ and $p_j$, and $d(r_i, r_j)$ denotes the distance between $r_i$ and $r_j$.  

\subsection{Reachability}
\label{ss:reachability}
The notion of reachability is needed to ensure realism in perturbed trajectories.  We begin by identifying the subset of POIs that can be reached from any particular POI, based on the relative physical distance between them.  
A threshold, $\theta$, represents the maximum distance that one can travel in a certain time period.  
$\theta$ can be specified directly, or be a function of $g_t$ and a given travel speed.

\begin{table}[t]
\centering
\caption{Commonly-Used Notation}
\label{tab:notation}
\begin{tabular}{P{1.2cm}p{6.5cm}}
\toprule
\textbf{Notation}       & \textbf{Meaning} \\
\midrule
$p, \mathcal{P}$        & Point of interest (POI), and set of POIs \\
$g_t, g_s$              & Time and space granularity \\ 
$t, T$                  & Timestep, and set of all timesteps \\
$\tau, \mathcal{T}$     & Trajectory, and set of trajectories \\
$\chi$                  & Dimension subscript (e.g., $t$ for temporal dimension) \\
$r_\chi, \mathcal{R}_\chi$ & STC region, and set of regions \\
$n$                     & Length of trajectory fragment (i.e., $n$-gram) \\
$w, \mathcal{W}^n$      & $n$-gram, and set of (reachable) $n$-grams \\
$\tau(a,b)$             & Trajectory fragment; covers $a$\textsuperscript{th} to $b$\textsuperscript{th} elements of $\tau$\\
$\theta$                & Reachability threshold\\
$d_\chi(\cdot)$         & Distance function for dimension $\chi$\\
$\epsilon$              & Privacy budget \\
\bottomrule
\end{tabular}
\end{table}

\begin{definition}[Reachability]
    \label{def:reachabilty}
    A POI $p_b \in \mathcal{P}$ is reachable from $p_a$ at time $t$ if $d_s(a,b) \leq \theta(t)$, where $p_a\in \mathcal{P}$.
\end{definition}
Reachability prevents illogical trajectories from being produced. For example, a trajectory of \{New York City, Tokyo, London\} would be unrealistic if the time granularity was one hour.  
Alternatively, imagine the trajectory: \{Grand Central Station, Empire State Building, Central Park\}.  A realistic bigram to perturb to might be \{Empire State Building, Times Square\} (i.e., the two locations can be reached within one hour), whereas \{Empire State Building, Mount Rushmore\} is unrealistic as it would not satisfy the reachability constraint.
The definition of reachability accommodates time-varying and asymmetric distances (e.g., caused by congestion and one-way roads, respectively).
For $n$-gram perturbations, $\mathcal{W}^n$ is the set of all $n$-grams that satisfy the requirements of reachability.
Formally, for the $n$-gram $w = \{p_a, ... p_i, p_{i+1}, ..., p_b\}$, the reachability constraint requires $p_{i+1}$ to be reachable from $p_i$ at time $t_i$ for all $a \leq i < b$.

\subsection{Privacy Mechanisms}
\label{ss:privacy-metrics}

\begin{definition}[$\epsilon$-local differential privacy \cite{Duchi2013}]
A randomized mechanism $\mathcal{M}$ is $\epsilon$-local differentially private if, for any two inputs $x, x'$ and output $y$:
\begin{equation}
\textstyle
    \frac{\Pr[\mathcal{M}(x) = y]}{\Pr[\mathcal{M}(x') = y]} \leq e^\epsilon
    \label{eq:ldp}
\end{equation}
\end{definition}
Whereas centralized DP allows the aggregator to add noise, LDP ensures that noise is added to data before it is shared with an aggregator. 
Like its centralized analogue, LDP possesses two fundamental properties that we use in our mechanism \cite{Cormode2018b}.  
The first is the composition theorem, which states that one can apply $k$ $\epsilon_i$-LDP mechanisms, with the result satisfying $\epsilon$-LDP, where $\epsilon = \sum_i \epsilon_i$.  The second property allows post-processing to be performed on private outputs without affecting the privacy guarantee.

\begin{definition}[Exponential Mechanism (EM) \cite{McSherry2007}]
For an input $x$ and output $y \in \mathcal{Y}$, the result of mechanism $\mathcal{M}$ is $\epsilon$-differentially private if one randomly selects $y$ such that:
\begin{equation}
\textstyle
    \Pr[\mathcal{M}(x) = y] = \frac{\exp\left(\epsilon q(x,y)/2\Delta_q\right)}{\sum_{y_i\in \mathcal{Y}}\exp\left(\epsilon q(x,y_i)/2\Delta_q\right)}
    \label{eq:exp-mech}
\end{equation}
where, $q(x,y)$ is a quality (or utility) function, and $\Delta_q$ is the sensitivity of the quality function, defined as $\max_{y, y'}|q(x,y) - q(x, y')|$.  
\end{definition}
The utility of the EM can be written as:
\begin{equation}
    \textstyle
    \Pr\left[q(x, y) \leq OPT_q - \frac{2\Delta_{q}}{\epsilon}\left(\ln{\frac{|\mathcal{Y}|}{|\mathcal{Y}_{OPT}|} + \zeta}\right)\right] \leq e^{-\zeta}
    \label{eq:em-utility}
\end{equation}
where $OPT_q$ is the maximum value of $q(x, y)$, and $\mathcal{Y}_{OPT} \subseteq \mathcal{Y}$ is the set of outputs where $q(x, y) = OPT_q$ \cite{Dwork2014}.

Although more commonly used for centralized DP, the EM can be applied in LDP, where we consider different inputs (for LDP) to be equivalent to neighboring datasets of size 1 (for centralized DP).
By using the EM with a distance-based quality function, we achieve $\epsilon$-LDP (i.e., the probability ratio for any perturbation is upper-bounded by $e^\epsilon$).  This is in contrast to relaxations of LDP, such as metric-LDP \cite{Alvim2018}, context-aware LDP \cite{Acharya2019}, or input-discriminative LDP \cite{Gu2019}.  All of these have upper bounds for the probability ratio of a perturbation of the form $e^{f(x, x')}$, where $f$ is a function that quantifies the distance between two inputs $x$ and $x'$.  Note that our setting is a specific (stricter) case in which $f(x, x') = \epsilon$ for all $x, x'$.

\section{Trajectory Perturbation}
\label{s:solution}
In this section, we first present a global solution that perturbs the whole trajectory (Section \ref{ss:global-solution}), before outlining our $n$-gram-based mechanism that addresses the infeasible aspects of the global solution (Sections \ref{ss:main-solution-overview}--\ref{ss:poi-reconstruct}).
We provide theoretical analysis (Sections \ref{ss:privacy-analysis} and \ref{ss:efficiency-discussion}), a number of alternative approaches (Section \ref{ss:alternative-approaches}), and an outline of our multi-attributed distance function (Section \ref{ss:distance-function}).

\subsection{Global Solution}
\label{ss:global-solution}

\label{sss:global-solution-overview}
In the global solution, we model entire trajectories as points in high-dimensional space.  
Having instantiated all possible trajectories, we determine the distance between these high-dimensional points and the real trajectory, and use this distance to determine the probability distribution. 
The probability of $\tau$ being perturbed to $\hat{\tau}$ is:
\begin{equation}
    \label{eq:global-mechanism}
    \textstyle
    \Pr(\hat{\tau} = \tau_i) = \frac{\exp\left(-\epsilon d_\tau(\tau, \tau_i)/2\Delta_{d_\tau}\right)}{\sum_{\tau_i\in \mathcal{S}}\exp\left(-\epsilon d_\tau(\tau, \tau_i)/2\Delta_{d_\tau}\right)}
\end{equation}
where, $\mathcal{S}$ is the set of all possible trajectories, and $d_\tau$ is the distance function that represents the distance between trajectories (see Section \ref{ss:distance-function}).
We use the EM \eqref{eq:exp-mech} to perturb trajectories.
Proof that the global solution satisfies $\epsilon$-LDP follows from \eqref{eq:global-mechanism} and Definition 4.3.
\begin{theorem}
The utility of the global solution is expressed as:
\begin{equation}
\textstyle
    \Pr\left[d_\tau(\tau, \hat{\tau}) \leq -\frac{2\Delta_{d_\tau}}{\epsilon}\left(\ln{|\mathcal{S}| + \zeta}\right)\right] \leq e^{-\zeta}
    \label{eq:global-utility}
\end{equation}
\end{theorem}
\begin{proof}
The quality function is the distance function $d_\tau$, and its maximum value is obtained iff $\hat{\tau} = \tau$.  
Hence, $OPT_q = 0$ and $|\mathcal{Y}_{OPT}| = 1$.
Substituting this into \eqref{eq:em-utility} yields \eqref{eq:global-utility}.
\end{proof}

\label{sss:global-solution-analysis}
To assess the feasibility of the global solution, we first analyze its worst-case time complexity.
The number of possible timestep sequences is 
$\frac{|T|!}{|\tau|! \times (|T|-|\tau|)!}$
where $|T|$ is the number of timesteps, which is a function of $g_t$ (in minutes): $|T| = \frac{24\times60}{g_t}$.  
The number of possible POI sequences is $|\mathcal{P}|^{|\tau|}$.
Hence, the maximum size of $\mathcal{S}$ is:
$|\mathcal{S}| = \frac{|\mathcal{P}|^{|\tau|} \times |T|!}{|\tau|! \times (|T|-|\tau|)!}$
The global solution requires instantiating all trajectories in $\mathcal{S}$, the size of which grows exponentially with $|\tau|$. 

In reality, the reachability constraint reduces the number of possible trajectories.
Assuming that (on average) $\mu$\% of all POIs are reachable between successive timesteps, the number of possible trajectories is reduced by a factor of $\mu^{|\tau| - 1}$.  To illustrate this, imagine a small-scale example where $|\tau| = 5$, $g_t = 15$ minutes, $|\mathcal{P}| = 1{,}000$, and $\mu = 20\%$.  
Even under these settings, $|\mathcal{S}| \approx$ \num{9.78e19}, which means $\mathcal{S}$ remains computationally infeasible to compute and store. 

Variants of the EM approach have been proposed, with the aim of tackling the high cost of enumerating all possible outputs. 
The subsampled exponential mechanism applies the EM to a sample of possible outputs~\cite{Lantz2015}. 
In our case, we would need the sampling rate to be very small to make this approach computationally feasible.  
However, the highly skewed distribution of $d_{\tau}$ means that the sampling rarely selects trajectories with low $d_{\tau}$ values, which ultimately leads to poor utility in the perturbed dataset. 
The `Permute and Flip' approach \cite{McKenna2020} instead considers each output in a random order, and performs a Bernoulli test to see if it can be output.  
However, the success probability, which is proportional to $\exp(-\epsilon d_{\tau})$, is generally low, meaning that efficiency gains are limited. 
Instead, we look to alternative models to make our approach feasible. 

\subsection{\textit{n}-gram Solution Overview}
\label{ss:main-solution-overview}
Instead of considering trajectories as high-dimensional points, we propose using overlapping fragments of the trajectories to capture spatio-temporal patterns with efficient privacy-preserving computations.
Specifically, we consider a hierarchical $n$-gram-based solution that aims to privately perturb trajectories more quickly.
Using (overlapping) $n$-grams allows us to consider the spatio-temporal link between any $n$ consecutive points, which is necessary to accurately model trajectory data.  

\begin{figure}[t]
    \centering
    \includegraphics[width = .85\columnwidth]{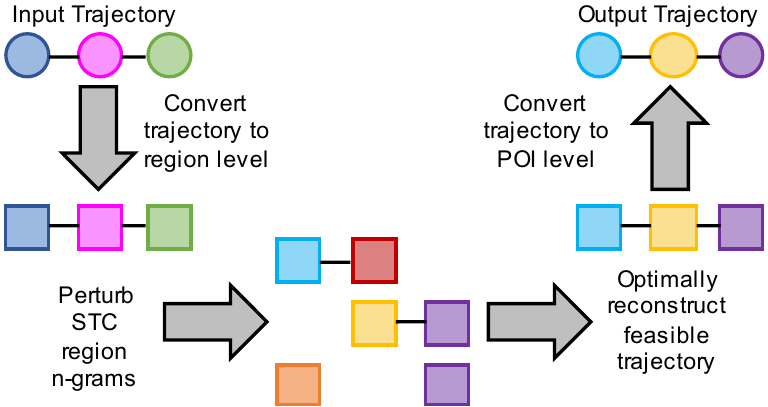}
    \caption{Solution Overview}
    \label{fig:mechanism-overview}
\end{figure}

Our solution (summarized in Figure \ref{fig:mechanism-overview}) has four main steps: hierarchical decomposition, $n$-gram perturbation, optimal STC region-level reconstruction, and POI-level trajectory reconstruction.  
Hierarchical decomposition is a pre-processing step that only uses public information, so it can be done a priori and without use of the privacy budget. 
Similarly, both trajectory reconstruction steps do not interact with private data, thus allowing us to invoke LDP's post-processing property  without using the privacy budget.

\subsection{Hierarchical Decomposition}
\label{ss:hierarchical-decomposition}
As $|\mathcal{P}|$ increases, the number of feasible outputs for POI-level perturbation grows exponentially.
To address this challenge, we utilize hierarchical decomposition to divide POIs into STC regions.  

\para{STC Region Composition}
We first divide the physical space into $R_{s}$ spatial regions.  
For each $r_{s} \in \mathcal{R}_{s}$, we create $R_{c}$ regions -- one for each POI category.  For each space-category region, we create $R_{t}$ regions, which represent coarse time intervals.  
POIs are then assigned to STC regions, based upon their location, opening hours, and category. 
POIs can appear in more than one STC region (e.g., they are open throughout the day and/or the POI has more than one category).
We remove all STC regions that have zero POIs within them (e.g., $r_{stc} = (\text{top of mountain}, \text{3am}, \text{church})$), which ensures that these regions will not be included in $\mathcal{W}^n$.  
As all this information is public, it does not consume any privacy budget. 

$\mathcal{R}_s$ can be formed using any spatial decomposition technique, such as uniform grids or clustering, or it can use known geography, such as census tracts, blocks, or boroughs.
We find that our mechanism is robust to the choice of spatial decomposition technique.
$\mathcal{R}_c$ can be derived from known POI classification hierarchies, such as those published by OpenStreetMap, etc.
$\mathcal{R}_t$ is most easily formed by considering a number of coarse (e.g., hourly) time intervals.

\para{STC Region Merging}
Depending on the number of STC regions, and the number of POIs within them, STC regions can be merged across any (or all) of the three dimensions.
For example, instead of $r_{stc} = $ (Main Street, 1am, Nightclub) and $r_{stc} = $ (Main Street, 11pm, Bar), they could be merged into $r_{stc} = $ (Main Street, 11pm-2am, Nightlife), which represents merging in the time and category dimensions.

Merging regions is done primarily for efficiency reasons as it prevents many semantically similar, but sparsely populated, regions from existing.
Additional POI-specific information (e.g. popularity) can be included into merging criteria to prevent significant negative utility effects.  
For example, if the data aggregator wishes to preserve large spatio-temporal hotspots, they will want to prevent merging very popular POIs with semantically similar but less popular POIs.  
For example, consider a conference center complex.  Although all conference halls are semantically similar, one hall might have a large trade show, whereas the others may have small meetings.  
It is important not to merge all halls in this case, as this might result in less accurate responses to downstream data mining tasks.
Hence, we require that each STC region has $\kappa$ POIs associated with it where $\kappa$ is a pre-defined function of POI attributes.  

\begin{figure}[tb]
    \begin{subfigure}[b]{0.3\columnwidth}
        \centering
        \includegraphics{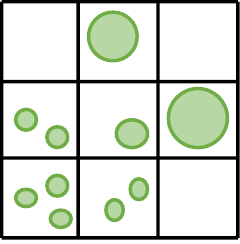}
        \caption{}
        \label{fig:merging-l}
    \end{subfigure}
    \hfill
    \begin{subfigure}[b]{0.3\columnwidth}
        \centering
        \includegraphics{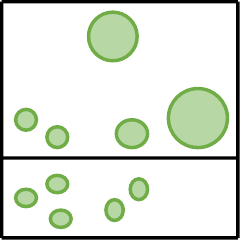}
        \caption{}
        \label{fig:merging-c}
    \end{subfigure}
    \hfill
    \begin{subfigure}[b]{0.3\columnwidth}
        \centering
        \includegraphics{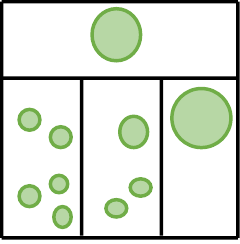}
        \caption{}
        \label{fig:merging-r}
    \end{subfigure}
    \caption{Illustrative example of the STC region merging}
    \label{fig:mergin-example}
\end{figure}

To further illustrate this, consider Figure \ref{fig:merging-l}, which shows ten POIs, divided in a $3\times3$ grid where larger circles indicate a more popular POI.  
Figure \ref{fig:merging-c} shows how regions might be merged if we only consider geographic proximity, whereas Figure \ref{fig:merging-r} shows the resultant regions when merging accounts for perceived popularity.  
We see more POIs in regions with less popular POIs, whereas very popular POIs exist singly in a region.
Deciding along which dimensions to merge regions, as well as the priority and extent of merging, depends ultimately on the utility goals of the data aggregator.  
For example, if preserving the category of POIs is important, merging in the time and space dimensions first would be preferable.  

\para{$n$-gram Set Formation}
As a final pre-processing step, we define $\mathcal{W}^n$ by first instantiating all possible $n$-gram combinations of STC regions.  We then remove all $n$-gram combinations that do not satisfy the reachability constraint.  For the region level, we deem any $r_a$ and $r_b$ to be reachable if there is at least one $p_i \in r_a$ and at least one $p_j \in r_b$ that satisfy reachability.

\subsection{\textit{n}-gram Perturbation}
\label{ss:ngram-perturbation}

Once $\mathcal{W}^n$ has been defined, we convert each trajectory from a sequence of POI-timestep pairs to a sequence of STC regions.
The next step is to perturb the STC regions of $\tau$
by using overlapping $n$-grams and the EM.  

\para{Notation}
We define $Z$ to be a set that holds all the perturbed $n$-grams of $\tau$.
We then use $z(a,b) = \{\hat{r}_a,..., \hat{r}_i,... \hat{r}_b\}$ to be the perturbed $n$-gram, where $a$ and $b$ are the indices of $\tau$ and $z(a,b) \in Z$.  
Importantly, there is a subtle difference between $\hat{\tau}(a,b)$ and $z(a,b)$. 
In $Z$, for any timestep, there are multiple possible regions associated with each trajectory point, whereas $\hat{\tau}$ is the final reconstructed trajectory and so there is only one region for each trajectory point.

\begin{figure}[tb]
    \centering
    \includegraphics[width = \columnwidth]{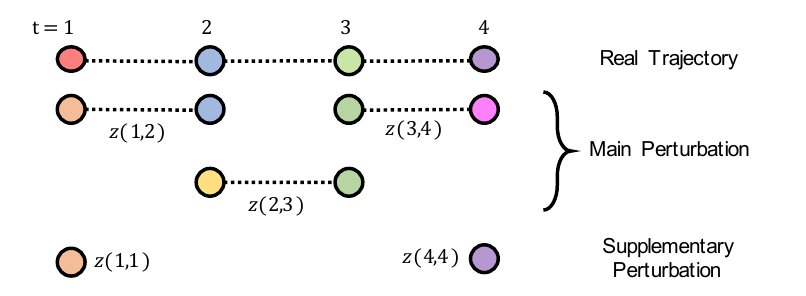}
    \caption{Main and supplementary perturbation mechanisms; different colors indicate different STC regions}
    \label{fig:perturbation}
\end{figure}

\para{Main Perturbation}
For each perturbation, we take $\mathcal{W}^n$ and define the probability that $\tau(a,b)$ is perturbed to $w_i \in \mathcal{W}^n$ as:
\begin{equation}
\textstyle
    \label{eq:ngram-perturbation}
    \Pr(z(a,b) = w_i) = \frac{\exp\left(-\epsilon' d_w(\tau(a,b), w_i)/2\Delta_{d_w}\right)}{\sum_{w\in \mathcal{W}^n}\exp\left(-\epsilon' d_w(\tau(a,b), w_i)/2\Delta_{d_w}\right)}
\end{equation}
where $\epsilon' = \frac{\epsilon}{|\tau| + n - 1}$, and $d_w(\tau(a,b), w_i)$ is the function that quantifies the distance between $n$-grams.  
To ensure that $n$-grams are perturbed, we specify the ranges of $a$ and $b$ such that $a = (1, |\tau|-n+1)$ and $b = (n, |\tau|)$.
Once these probabilities have been defined, we use the EM to sample from $\mathcal{W}^n$ and we store $z(a,b)$ in $Z$.  We repeat this for increasing values of $a$ and $b$ (see Figure \ref{fig:perturbation}).

Using overlapping $n$-grams gives better accuracy than using non-overlapping $n$-grams or merely perturbing points independently.  
It lets us repeatedly `query' a point multiple times, meaning that we gather more information about it while continuing to guarantee LDP. 
Overlapping $n$-grams simultaneously allows us to query a larger portion of the entire trajectory, which enables us to base each perturbation upon a wider range of semantic information.
For example, $\hat{\tau}(3)$ is determined based on information from $\tau(1,3)$, $\tau(2,4)$, and $\tau(3,5)$, assuming $n=3$. Hence,  $\tau(3)$ is `queried' $n$ times, neighboring points $n-1$ times, etc., in addition to using information from $2n-1$ points to influence the perturbation of $\tau(3)$.

\para{End Effects}
When $n\geq2$, the start and end regions in a trajectory are not covered $n$ times.  
For example, when $|\tau| = 4$ and $n = 2$, the main perturbation step covers the first and last regions once only (with $z(1,2)$ and $z(3,4)$, respectively).  
To ensure that all timesteps have the same number of perturbed regions, we conduct extra  perturbations with smaller $n$-grams. 
This supplementary perturbation is performed in the same manner as \eqref{eq:ngram-perturbation}, but with different $\mathcal{W}^n$ sets and different bounds for $a$ and $b$.  
In our example, we would obtain $z(1,1)$ and $z(4,4)$ using $\mathcal{W}^1$ (as illustrated in Figure \ref{fig:perturbation}). 

\para{Theoretical Utility}
As the solution has multiple post-processing stages, a theoretical utility guarantee for the entire mechanism remains elusive. However, we can analyze the utility of the $n$-gram perturbation step.
\begin{theorem}
The utility of the $n$-gram perturbation stage is:
\begin{equation}
\textstyle
    \Pr\left[d_w(\tau(a,b), w) \leq -\frac{2\Delta_{d_w}}{\epsilon'}\left(\ln{|\mathcal{W}^n| + \zeta}\right)\right] \leq e^{-\zeta}
    \label{eq:ngram-utility}
\end{equation}
\end{theorem}
\begin{proof}
The quality function is the distance function $d_w$, and its maximum value is obtained iff $z(a,b) = \tau(a,b)$.  
Hence, $OPT_{d_w} = 0$ and $|\mathcal{\mathcal{W}}^n_{OPT}| = 1$.
Substituting this into \eqref{eq:em-utility} yields \eqref{eq:ngram-utility}.
\end{proof}
Hence, the utility is dependent on the size of the feasible $n$-gram set, which itself is influenced by $n$, the granularity of hierarchical decomposition, and the strictness of the reachability constraint.
Utility is also affected by trajectory length, as $\epsilon'$ is a function of $|\tau|$.

\subsection{Region-Level Trajectory Reconstruction}
\label{ss:opt-reconstruction}
\begin{figure}[tb]
    \centering
    \begin{subfigure}[b]{0.35\columnwidth}
        \centering
        \includegraphics{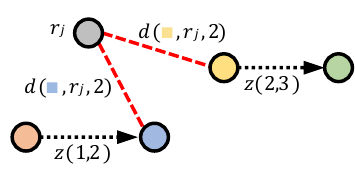}
        \caption{Node Error}
        \label{fig:real}
    \end{subfigure}
    \hfill
    \begin{subfigure}[b]{0.5\columnwidth}
        \centering
        \includegraphics{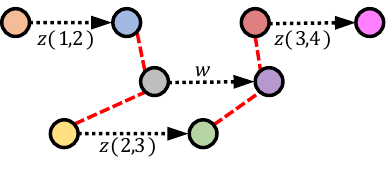}
        \caption{Bigram Error}
        \label{fig:kde}
    \end{subfigure}
    \caption{Illustrative example of error terms: error equals sum of red dashed line distances}
    \label{fig:error-terms}
\end{figure}

Given a collection of perturbed $n$-grams, we define an optimization problem that reconstructs a trajectory between points in $\hat{\tau}$ and the perturbed $n$-grams in $Z$.
This is post-processing of the LDP output, and does not consume any privacy budget. 
We define two error terms (illustrated in Figure \ref{fig:error-terms}) that measure the similarity of regions to the perturbed data. 
By perturbing each point in $\tau$ $n$ times (by using overlapping $n$-grams), we magnify this effect.  

The first error term is the `region error' $e(r_j, i)$, which is the distance between $r_j$ and the perturbed $n$-grams in location $i$ in $Z$. 
It is defined as:
\begin{equation}
\textstyle
    \label{eq:region-error}
    e(\hat{r}_j, i) = \sum d(\hat{r}_j, y_i)
\end{equation}
where $y_i$ is the region from $z(a,b) \in Z$ iff $a \leq i \leq b$, where $a$, $b$, and $i$ are trajectory indices.  
The second error term is the `bigram error', $e(i,w)$, which is the sum of the two relevant region error terms. 
More formally, it is defined as:
\begin{equation}
\textstyle
    \label{eq:bigram-error}
    e(i, w) = e\left(w(1), i\right) + e\left(w(2), i+1\right)
\end{equation}
where $w$ is a region-level bigram in $\mathcal{W}^2$, with $w(1)$ and $w(2)$ being the first and second regions in $w$, respectively.  
 
We now define the minimization problem as:
\begin{equation}
\textstyle
    \label{eq:objective-function}
    \min \; \sum^{|\tau|-1}_{i=1} x^w_i e(i, w)
\end{equation}
%
\begin{equation}
\textstyle
    \label{eq:continuity-1}
    \text{s.t.} \quad x^w_i \cdot q(w_i, w_{i+1}) = x^w_{i+1} \cdot q(w_i, w_{i+1}) \quad \forall \; 1 \leq i < |\tau|
\end{equation}
\begin{equation}
\textstyle
    \label{eq:continuity-2}
   q(w_i, w_{i+1}) = 
        \begin{cases}
            1  & \text{if } w_i(2) = w_{i+1}(1)\\
            0  & \text{otherwise}
        \end{cases}
\end{equation}
\begin{equation}
\textstyle
    \label{eq:total-bigrams}
    \sum_{i=1}^{|\tau|-1} x^w_i = |\tau| - 1
\end{equation}
\begin{equation}
\textstyle
    \label{eq:each-has-1}
    \sum_{w\in\mathcal{W}^2} x^w_i = 1 \quad \forall \; 1 \leq i < |\tau|
\end{equation}
where, $x^w_i$ is a binary variable encoding whether $w$ is selected for index $i$.
The objective \eqref{eq:objective-function} is to minimize the total bigram error across the trajectory.  
\eqref{eq:continuity-1} and \eqref{eq:continuity-2} are continuity constraints that ensure consecutive bigrams share a common region.
\eqref{eq:total-bigrams} ensures that the number of bigrams selected is correct, and \eqref{eq:each-has-1} ensures only one bigram is associated with each point in the trajectory.

\para{Efficiency Discussion}
Assuming that the space, time, and category granularities are well-chosen such that $|\mathcal{W}^2| \ll |\mathcal{R}|^2$, the scale of the optimization problem will generally be within the scope of most linear programming solvers (see Section \ref{ss:efficiency-discussion}).  
Nevertheless, we introduce a step to further limit the set of possible bigrams that can appear in the reconstructed trajectory.
Once $n$-gram perturbation is complete, we obtain the minimum bounding rectangle (MBR) defined by all $r_{stc} \in Z$. 
From this, we define $\mathcal{P}_{mbr} \subseteq \mathcal{P}$, which contains all the POIs in this MBR, and $\mathcal{R}_{mbr}$, which is the set of STC regions that contain at least one POI in $\mathcal{P}_{mbr}$.  
From this, we define $\mathcal{W}^2_{mbr}$ as the set of feasible bigrams formed from $\mathcal{R}_{mbr}$, and we use this set in the reconstruction. 
Performing this step does not prevent the optimal reconstructed trajectory from being found, as the reconstruction seeks to minimize the error with respect to the perturbed $n$-grams in $Z$, all of which are included in $\mathcal{R}_{mbr}$.

\subsection{POI-Level Trajectory Reconstruction}
\label{ss:poi-reconstruct}
The final step is to express the output trajectories in the same format as the input trajectories.
Whereas converting a trajectory from POIs to STC regions is relatively straightforward, the converse operation is non-trivial as there can be many possible POI-level trajectories corresponding to a certain sequence of STC regions. Furthermore, the reachability requirement means that some trajectories are infeasible, and should not be published.  

We expect most POI-level trajectories to be feasible as $\mathcal{W}^n$ is defined based on the reachability criterion.  
Hence, we generate an individual trajectory at random and check that it satisfies the reachability constraint.  
If it does, we output it; if not, we generate another trajectory.  
We continue this until we generate a feasible trajectory, reach a threshold ($\gamma$), or exhaust all possible combinations.  
Experimentally, the threshold of $\gamma = 50{,}000$ was rarely reached.

When this trajectory sampling fails, it implies that the perturbed region sequence does not correspond to a feasible trajectory. 
If so, we randomly select a POI and time sequence and `smooth' the times such that they become feasible.  
For example, consider the region-level trajectory: \{(Restaurant, 9-10pm, Downtown), (Bar, 9-10pm, Downtown), (Bar, 9-10pm, Suburb)\}.  
Reachability may mean that the suburban bar is only reachable from the downtown bar in 55 minutes, meaning that it is impossible to visit all three venues in an hour.  
Accordingly, we smooth the timesteps such that either $\tau(1)$ occurs between 8 and 9pm, or $\tau(3)$ occurs between 10 and 11pm.

\subsection{Privacy Analysis}
\label{ss:privacy-analysis}
We now analyze the privacy of the $n$-gram-based solution through a sketch proof, and discuss some possible adversarial attacks.
\begin{theorem}
The perturbation of trajectory $\tau$ satisfies $\epsilon$-LDP.
\end{theorem}
\begin{proof}
Each $n$-gram $\tau(a,b)$  -- where $a = (1, |\tau|-n+1)$ and $b = (n, |\tau|)$ -- 
is perturbed with privacy budget $\epsilon'$ using \eqref{eq:ngram-perturbation}, and the EM.
This means each perturbation satisfies $\epsilon'$-DP, and there are $|\tau|-n+1$ of these perturbations.
Because of end effects, there are an additional $2(n-1)$ perturbations, each of which also satisfy $\epsilon'$-DP.
Using sequential composition, the resultant output satisfies $(|\tau| + n - 1)\epsilon'$-DP.
As $\epsilon' = \frac{\epsilon}{(|\tau| + n - 1)}$, the overall mechanism therefore satisfies $\epsilon$-DP.
As the size of adjacent datasets is 1, $\epsilon$-DP results are equivalent to $\epsilon$-LDP results.
\end{proof}

As we use publicly available external knowledge, we assume that an adversary has access to all the same knowledge.
However, external knowledge is \textit{only} used to enhance utility, whereas privacy is provided through the application of the EM (which could be done with no external knowledge).
Hence, an adversary cannot use this information to learn meaningful information with high probability.
As our solution is predicated on a `one user, one trajectory' basis, inference attacks based on repeated journeys from the same user are prevented by definition.
We can use sequential composition to extend our solution to the multiple release setting; assuming each of $k$ trajectories is assigned a privacy budget of $\epsilon$, the resultant release provides $(k\epsilon)$-LDP to each user.
Finally, unlike in other works that consider continuous data sharing \cite[e.g.][]{Acs2014, Dwork2010, Cao2017, Kellaris2014}, our setting sees the user share all data at the end of their trajectory.
Hence, as we provide user-level $\epsilon$-LDP, the LDP privacy guarantee protects against spatial and temporal correlation attacks.

\subsection{Computational Cost}
\label{ss:efficiency-discussion}

We now discuss the computational costs of the proposed $n$-gram-based approach, which highlight how it is highly practical.  
Also, as each perturbation can be done locally on a user's device, the entire data collection operation is inherently distributed and scalable.

\para{Choice of $n$}
Precomputing $\mathcal{W}^n$ requires $\mathcal{O}(|\mathcal{P}|^n)$ space, which becomes infeasible for large cities and $n \geq 3$.  
If $n \geq 3$ and $|\mathcal{P}|$ is large, feasible $n$-grams can be computed `on-the-fly', although this attracts a significant runtime cost.  
While these effects are partially mitigated by using STC regions, we recommend choosing $n=2$ (bigrams) as $n \geq 3$ will be unrealistic in most practical settings.

\para{Time Complexities}
Converting a trajectory from the POI-level to STC region level has time complexity $\mathcal{O}(|\tau|)$ and the perturbation phase has time complexity $\mathcal{O}(n|\tau|)$. 
Converting trajectories from the STC region level back to the POI level (assuming time smoothing does not need to be performed) has a worst-case time complexity of $\mathcal{O}(\gamma(|\tau|+|\tau-1|))$.  
This is because $|\tau|$ POIs need to be selected, and then reachability checks need to be performed on each link (of which there are $|\tau|-1$ in total).  
In the worst-case, this process is repeated $\gamma$ times, hence $\mathcal{O}(\gamma(|\tau|+|\tau-1|))$.
We find that time smoothing is needed for around 2\% of trajectories on average, and so we devote little focus to its runtime effects here.

\para{Optimal Reconstruction Complexity}
Section \ref{ss:opt-reconstruction} presents the optimal reconstruction phase, which uses integer linear programming.
Here, we briefly discuss the scale of the problem in terms of the number of variables and constraints.  
Let $\phi = |\mathcal{R}_{mbr}|$.  
The number of feasible bigrams will be $\phi^2$ in the worst case, although the reachability constraint reduces this in practice.
From the definition of $x^w_i$, we see that there are $\phi(|\tau|-1)$ variables (i.e., one $x^w_i$ per bigram, per trajectory point).
The continuity constraints -- \eqref{eq:continuity-1} and \eqref{eq:continuity-2} -- impart $\phi(|\tau|-1)$ constraints, and the capacity constraints -- \eqref{eq:total-bigrams} and \eqref{eq:each-has-1} -- impart $|\tau|-1$ constraints.  
Hence, the optimization problem has $(\phi|\tau| + |\tau| - \phi -1)$ constraints in total.
Closed-form expressions for the expected runtime of optimization problems depend on the exact solver chosen, and are typically fast in practice \cite[see][]{vandenBrand2020, Cohen2020}.

\subsection{Alternative Approaches}
\label{ss:alternative-approaches}
We compare our solution to other approaches, summarized here. 

\para{Using Physical Distance Only}
The most basic distance-based perturbation mechanism (called \textsc{PhysDist}) would ignore external knowledge and only use the physical distance between POIs/regions. 

\para{POI-level n-Gram Perturbation}
Our mechanism can be applied just on the POI-level.  
\textsc{NGramNoH} perturbs the time and POI dimensions separately in order to control the size of $\mathcal{W}^n$.  This requires splitting the privacy budget more (i.e., $\epsilon' = \frac{\epsilon}{2|\tau| + n - 1}$).

\para{Independent POI Perturbation}
The simplest approach is to perturb each POI independently of all others.  
We consider two variations of this approach: one where the reachability constraint is considered during perturbation (\textsc{IndReach}), and one where it is not (\textsc{IndNoReach}).  
To ensure that feasible trajectories are output when using \textsc{IndNoReach}, we use post-processing to shift the perturbed timesteps to ensure a `realistic' output.
While such methods make less intensive use of the privacy budget, they fail to account for the intrinsic relationship between consecutive points.
However, when temporal gaps between POIs are large, the reachability constraint becomes less influential, making these methods more attractive. 

\para{Other LDP Relaxations}
As discussed in Section \ref{s:related-work}, a number of distance-based perturbation mechanisms that were inspired by the principles of (L)DP exist.  However, although these approaches possess their own theoretical guarantees, they do not satisfy $\epsilon$-LDP, which makes them incomparable with our mechanism.

\subsection{Distance Function}
\label{ss:distance-function}

\label{sss:distance-definitions}
We now outline the semantic distance functions used throughout our work.
Note that our mechanism is not reliant on any specific distance/quality function -- any other distance function can be used, without needing to change the mechanism.

\para{Physical Distance}
\label{sss:physical-distance}
We use $d_s(p_a, p_b, t)$ to denote the physical distance from $p_a$ to $p_b$ at time $t$, which can be derived using any distance measure (e.g., Euclidean, Haversine, road network).  To get the distances between STC regions, we obtain the distance between the centroids of the POIs in the two regions.  
We similarly use $d_s(r_a, r_b)$ to denote the physical distance between $r_a$ and $r_b$.

\para{Time Distance}
\label{sss:time-distance}
The time distance between regions is defined as the absolute time difference between two STC regions. That is, $d_t(r_a, r_b) = |t_a - t_b|$.
We limit time distances to ensure that no time distance is greater than 12 hours.  
Where STC regions are merged in the time dimension, we use the time difference between the centroids of the merged time intervals.  
For example, if two regions cover 2-4pm and 5-7pm, $d_t = |3 - 6| = 3$ hours.

\begin{figure}[t]
    \centering
    \includegraphics{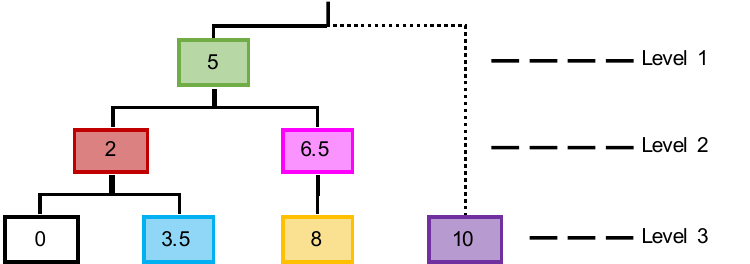}
    \caption{$\bm{d_c}$ values, relative to left-most level 3 node}
    \label{fig:category-distances}
\end{figure}

\para{Category Distance}
\label{sss:cat-distance}
Category distance, $d_c$, is quantified using a multi-level hierarchy. (We use three, although any number of levels can be used.)
Figure \ref{fig:category-distances} illustrates how $d_c$ varies across hierarchical levels, relative to the leftmost level 3 (white) node.  We define category distance to be symmetric (i.e., $d_c(\text{Shoe Shop}, \text{Shopping}) = d_c(\text{Shopping}, \text{Shoe Shop})$.  
If two POIs or regions do not share a level 1 category, 
we deem them to be unrelated and $d_c = 10$ (indicated by the dotted line and purple node in Figure \ref{fig:category-distances}).

\para{Combining Distances}
\label{sss:combining-distances}
Distance functions are combined as follows:
\begin{equation}
\textstyle
    d(r_a, r_b) = \left(d_s(r_a, r_b)^2 + d_t(r_a, r_b)^2 + d_c(r_a, r_b)^2 \right)^{1/2}
\end{equation}
To determine the `distance' between two $n$-grams, we use elementwise summation.  For example, the distance between two bigrams -- $w_i = \{i_1, i_2\}$ and $w_j = \{j_1, j_2\}$ -- is calculated as $d(i_1, j_1)$ + $d(i_2, j_2)$.  More generally, if $w_1$ and $w_2$ are two $n$-grams: 
\begin{equation}
    \label{eq:ngram}
    \textstyle
    d_w(w_1, w_2) = \sum^n_{a=1} d(i_a, j_a)
\end{equation}

\section{Experimental Set-Up}
\label{s:expts}
Our experiments seek to: 
\begin{enumerate*}[label={\alph*)}]
\item analyze our mechanism and gather insights from its behavior;
\item compare our approach to comparable alternatives; and
\item demonstrate the practical utility of our mechanism in the context of application-inspired queries.
\end{enumerate*}

\subsection{Data}
\label{ss:data}
We use a range of real, synthetic, and semi-synthetic trajectory datasets. 
There is a chronic lack of high quality, publicly available POI sequence data.
Hence, we augment existing real datasets to make them suitable for a comprehensive evaluation. 

\subsubsection{Real Data}
\label{sss:real-data}
We combine Foursquare check-in \cite{Yang2014} and historic taxi trip \cite{NYC2013} data, both from New York City.
The set of POIs is taken from all POIs that appear in the raw Foursquare dataset, from which we take the $|\mathcal{P}|$ most popular as our set $\mathcal{P}$.
We concatenate the pick-up and drop-off locations of each taxi driver's daily trips in order to protect their business-sensitive movements. 
We match the co-ordinate data with the nearest POI.
If no POIs within 100m are found, we discard the point. 
We clean the data by removing repeat points with the same venue ID or exact latitude-longitude location.  
Where points occur less than $g_t$ minutes apart, we remove all but one point at random.  
For category information, we use the publicly available Foursquare category hierarchy \cite{FoursquareHierarchy2020} to assign a single category to each POI.  
We manually specify opening hours for each broad category (e.g., `Food', `Arts and Entertainment'), and set all POIs of that (parent) category to have those hours.  
However, the mechanism is designed to allow POI-specific opening hours.

\subsubsection{Semi-Synthetic Data}
\label{sss:safegraph-data}
We use Safegraph's Patterns and Places data \cite{Safegraph2020} to semi-synthetically generate trajectories.  
We randomly determine $|\tau|$ using a uniform distribution with bounds (3,8), and the start time using a uniform distribution with bounds (6am, 10pm).  
The starting POI is selected at random from the popularity distributions of the day/time in question.  
We sample from the distribution of dwell times at each POI to determine the time spent in one location, and we sample the time spent traveling to the next POI uniformly from (0, 60) minutes.  
The next POI is sampled at random, based on the popularity distribution at the expected arrival time (based on the POIs that are `reachable').
This process continues until a trajectory is generated.  
Safegraph uses the NAICS hierarchy \cite{NAICS2020}, and we use this system for the category hierarchy.
Opening hours information for POIs is sparse, and so we manually assign general opening hours to categories, as in the Taxi-Foursquare data.

\subsubsection{Campus Data}
\label{sss:synthetic-data-campus}
We generate trajectories based on the University of British Columbia campus \cite{UBCcampus2016}.
The 262 campus buildings act as POIs, and nine POI categories exist, such as `academic building', `student residence', etc.  
We determine trajectory length and start time in the same way as for the Safegraph data.  
For each subsequent timestep in the trajectory, we sample from a uniform distribution with bounds ($g_t$, 120) minutes.  
The category of the first POI is chosen at random,
and the exact POI is chosen at random from all POIs in the selected category.  
For each subsequent POI, the POI is chosen from the set of reachable POIs based on the preceding POI, the time gap, and the time of day.
We artificially induce three popular events into the synthetic trajectories by picking a point in the trajectory, and controlling the time, POI, and category of the trajectory point.  
The remainder of the trajectory is generated as per the previously outlined method.  
The three popular events are: 500 people at Residence A at 8-10pm; 1000 people at Stadium A at 2-4pm; and 2000 people in some academic buildings at 9-11am.

\subsubsection{External Knowledge Specification}
\label{sss:external-knowledge}
Although we specify external knowledge manually, more scalable, operator-independent methods are possible.
For example, APIs of major location-based services (e.g., Google Maps) can be used to query thousands of POIs efficiently and cheaply.
In the case of Google Maps, information such as location, opening hours, category, price levels, etc. can be obtained directly through their API.
This information can be stored in the POI-level database, with which the mechanism interacts.

\subsubsection{Pre-Processing Costs}
\label{sss:pre-processing-costs}
\begin{figure*}
\centering
\parbox[t]{0.65\columnwidth}{\null
\centering
    \begin{subfigure}[b]{0.65\columnwidth}
    \centering
        \includegraphics[height = 0.4cm]{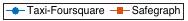}
    \end{subfigure}
    \\
    \begin{subfigure}[b]{0.3\columnwidth}
    \centering
        \includegraphics[height = 3cm]{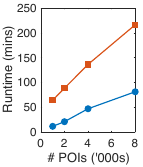}
    \end{subfigure}
    \hfill 
    \begin{subfigure}[b]{0.3\columnwidth}
    \centering
        \includegraphics[height = 3cm]{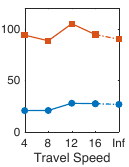}
    \end{subfigure}
  \captionof{figure}{Pre-processing runtime costs}
  \label{fig:preprocessing}
}
\hspace{0.5cm}
\parbox[t]{1.3\columnwidth}{\null
\centering
    \captionof{table}[t]{Mean NE between real and perturbed trajectory sets}
    \label{tab:traj-distances}  
    \begin{tabular}{c|ccc|ccc|ccc}
        \toprule
        \multirow{2}{*}{\textbf{Method}} & \multicolumn{3}{c|}{\textbf{Taxi-Foursquare}} & \multicolumn{3}{c|}{\textbf{Safegraph}} & \multicolumn{3}{c}{\textbf{Campus}} \\
        & $d_t$ & $d_c$ & $d_s$ 
        & $d_t$ & $d_c$ & $d_s$ 
        & {$d_t$} & {$d_c$} & {$d_s$} \\
        \midrule
        \textsc{IndNoReach} & 1.44 & 3.81 & 1.99 & 1.47 & 2.87 & 2.14 & 2.06 & 1.35 & 0.87 \\ 
        \textsc{IndReach} & 1.43 & 3.80 & 2.03 & 1.50 & 2.94 & 2.31 & 2.03 & 1.39 & 0.89 \\ 
        \textsc{PhysDist} & 1.61 & 8.74 & \textbf{1.85} & 1.62 & 8.38 & \textbf{2.10} & 2.16 & 3.04 & 0.90 \\ 
        \textsc{NGramNoH} & 1.63 & 4.25 & 2.07 & 1.62 & 3.37 & 2.33 & 2.14 & 1.46 & 0.88 \\ 
        \textsc{NGram} & \textbf{1.18} & \textbf{1.82} & 2.24 & \textbf{0.93} & \textbf{1.31} & 2.12 & \textbf{1.21} & \textbf{0.81} & \textbf{0.83} \\  
        \bottomrule
    \end{tabular}
}
\end{figure*}

The pre-processing necessary for our experiments is split into three parts: (a) POI processing, hierarchical decomposition, and region specification; (b) trajectory composition; and (c) trajectory filtering.
Part (a) is a one-time operation that creates the necessary data structures.
The impact of specifying external knowledge is negligible as the data structures (e.g., $\mathcal{R}$, $\mathcal{W}^n$) need to be created regardless.
Figure \ref{fig:preprocessing} shows the runtime costs for pre-processing step (a) for the two large-scale datasets.
The runtime is heavily dependent on the size of $\mathcal{P}$, but less influenced by the reachability constraint.
It is independent of other variables, such as trajectory length and privacy budget.
Although it is a one-time operation, localized updates can be performed to reflect changes in the real world (e.g., changes in POI opening times, new roads affecting reachable POIs).
Despite the large runtime, we argue that it is an acceptable cost, especially as it is a one-time process.
Parts (b) and (c) are only necessary as we are simulating the perturbation of thousands of trajectories.
In a practical setting, parts (b) and (c) are negligible as the trajectory data is created by each user and, by definition, a real trajectory should satisfy reachability and other feasibility constraints.
If there are infeasible aspects in a trajectory, smoothing operations can be performed in sub-second time.

\subsection{Experimental Settings}
\label{ss:expt-setup}
We set $g_t = 10$ minutes, $n = 2$ for all $n$-gram-based methods, and 
$|\mathcal{P}| = 2{,}000$.  
We set $\epsilon=5$, in line with other real-world deployments of LDP by Apple \cite{Apple2017} and Microsoft \cite{Ding2017}. 
We assume all travel is at 4km/hr (campus data) and 8km/hr (Taxi-Foursquare and Safegraph data). 
These speeds correspond to approximate walking and public transport speeds in cities, once waiting times, etc. have been included.
We consider the effects of varying these parameters in Section \ref{ss:param-variation}. 
We use Haversine distance throughout.  
We filter $\mathcal{T}$ to remove trajectories that do not satisfy the reachability constraint or where POIs are `visited' when they are closed.  
In general, the size of $\mathcal{T}$  (once filtered) is in the range of 5,000-10,000.

When creating STC regions, we divide the physical space using a $g_s \times g_s$ uniform grid.  
The finest granularity we consider is $g_s = 4$, and we use coarser granularities ($g_s = \{1, 2\})$ when performing spatial merging.  We use the first three levels of the Foursquare and NAICS category hierarchies (for the Taxi-Foursquare and Safegraph data, respectively), and use the category distance function outlined in Section \ref{ss:distance-function}.  Using these levels ensures that $d_c = 10$ for POIs with completely different categories.  STC regions have a default time granularity of one hour.  By default, we perform spatial merging first, followed by time merging, and category merging, and $\kappa = 10$.

We do not find any suitable alternatives in the literature---all existing work is based in the centralized DP domain, or uses relaxed definitions of LDP.  
Instead, we compare to the alternative approaches outlined in Section~\ref{ss:alternative-approaches}.

\subsection{Utility Measures}
\label{ss:utility}
We assess the accuracy of our perturbed trajectories through a range of measures.  First, we examine the distance between the real and perturbed trajectories.
We normalize the distance values by $|\tau|$ and use the term `normalized error' (NE) henceforth.  For this, we use the same distance definitions as outlined in Section \ref{ss:distance-function}.

\subsubsection{Preservation Range Queries}
We also examine a set of `preservation range queries' (PRQs).  
That is, for each point in each trajectory, we check to see whether the perturbed POI is within $\delta$ units of the true POI.  
For example, a location PRQ might examine whether $\hat{p}_i$ is within 50 meters of $p_i$.
We conduct PRQs in all three dimensions, and define the utility metric $PR_{\chi}$ as:
\begin{equation}
\textstyle
    PR_{\chi} = \frac{1}{\mathcal{|T|}}\sum_{\tau \in \mathcal{T}} \left(\frac{1}{|\tau|}\sum^{|\tau|}_{i = 1} \pi(p_i, \hat{p}_j, \delta_{\chi})\right) \times 100\%
\end{equation}
where  $\pi(p_i, \hat{p}_j, \delta)$ equals 1 if $d_{\chi}(p_i, \hat{p}_j) \leq \delta$, and zero otherwise.  
For time PRQs, $t_i$ and $\hat{t}_i$ replace $p_i$ and $\hat{p}_i$, respectively.

\subsubsection{Hotspot Preservation}
For each POI, spatial region, or category, we define a \textit{spatio-temporal hotspot} as the time interval during which the number of unique visitors is above a threshold $\eta$.  
A hotspot is characterized by $h = \{t_s, t_e, p_i, c\}$, where $t_s$ and $t_e$ are the start and end times, $p_i$ is the POI, and $c$ is the maximum count reached in the interval.  
Note that multiple hotspots can be associated with the same POI if popularity changes over time (e.g., a train station might have hotspots during the AM and PM peaks).  

We consider three spatial granularities: POI-level, and spatial regions defined by $4\times4$ and $2\times2$ grids, and $\eta = \{20, 20, 50\}$, respectively.  
We consider three category granularities (i.e., each hierarchical level), and $\eta = \{50, 30, 20\}$, for levels $\{1,2,3\}$ respectively.
We quantify hotspot preservation by calculating the `hotspot distance' between the hotspots in the perturbed and real data.  
If $\mathcal{H}$ and $\widehat{\mathcal{H}}$ are the hotspot sets in the real and perturbed data respectively, the average hotspot distance (AHD) between sets is: 
\begin{equation}
\textstyle
    \label{eq:hotspots}
    AHD(\mathcal{H}, \widehat{\mathcal{H}}) = \frac{1}{\left|\widehat{\mathcal{H}}\right|}\sum_{\hat{h}\in \widehat{\mathcal{H}}} \min_{h \in \mathcal{H}} \left(|t_s - \hat{t}_s| + |t_e - \hat{t}_e|\right)
\end{equation}
Note that, for each perturbed hotspot, we calculate the hotspot distance to each real hotspot (for the same space-category granularity) and report the minimum value.  This protects against cases in which there is not a one-to-one relationship between real and perturbed hotspots.  
We exclude hotspots in $\widehat{\mathcal{H}}$ for which there is no corresponding hotspot in $\mathcal{H}$.  We also record the absolute difference between $c$ and $\hat{c}$ for each hotspot pair.  When averaged across all hotspots, we obtain the average count difference, ACD.
\section{Results}
\label{s:results}
We compare our hierarchical solution to baseline methods in terms of NE and runtime.  
We also vary experimental and mechanism parameters, before evaluating on  application-inspired queries.  

\subsection{Baseline Comparison}
\label{ss:baseline-comparison}

\subsubsection{Normalized Error}
\label{sss:normalized-error}

Table \ref{tab:traj-distances} shows the distances between the real perturbed trajectories in all three dimensions.  
\textsc{NGram} is generally the best performing method across all datasets.
Comparison with \textsc{NGramNoH} demonstrates that a hierarchical approach provides accuracy benefits as well as efficiency benefits (as we will see). 
The importance of including external knowledge, such as category information, is emphasized when comparing performance with $\textsc{PhysDist}$, which performs worse than all other methods.  
Performance gains are primarily achieved in minimizing the category distance between real and perturbed trajectories.

\begin{table*}[t]
    \centering
    \caption{Average runtime in seconds; breakdown by main mechanism stages; values of 0.000s indicate runtimes that are less than 10\textsuperscript{-3}s; sum of individual runtime stages may not equal `Total' due to rounding}
    \label{tab:runtime-analysis}
    \begin{tabular}{c|ccccc|ccccc}
        \toprule
        \multirow{3}{*}{\textbf{Method}} & \multicolumn{5}{c|}{\textbf{Taxi-Foursquare}} & \multicolumn{5}{c}{\textbf{Safegraph}} \\
        & \multirow{2}{*}{\textbf{Perturb}} & \textbf{Reconst.} & \textbf{Optimal} & \multirow{2}{*}{\textbf{Other}} & \multirow{2}{*}{\textbf{Total}} & \multirow{2}{*}{\textbf{Perturb}} & \textbf{Reconst.} & \textbf{Optimal} & \multirow{2}{*}{\textbf{Other}} & \multirow{2}{*}{\textbf{Total}} \\
        & & \textbf{Prep.} & \textbf{Reconst.} & & & & \textbf{Prep.} & \textbf{Reconst.} & & \\
        \midrule
       \textsc{IndNoReach} & \textbf{0.005} & -- & -- & 0.714 & 0.720 & 0.006 & -- & -- & 0.786 & 0.791 \\ 
     \textsc{IndReach} & \textbf{0.005} & -- & -- & \textbf{0.000} & \textbf{0.006} & \textbf{0.005} & -- & -- & \textbf{0.000} & \textbf{0.006} \\ 
     \textsc{PhysDist} & 0.449 & 0.497 & 67.618 & \textbf{0.000} & 68.564 & 0.431 & 0.473 & 60.561 & \textbf{0.000} & 61.464 \\ 
     \textsc{NGramNoH} & 0.446 & 0.561 & 30.872 & \textbf{0.000} & 31.879 & 0.426 & 0.509 & 24.389 & \textbf{0.000} & 25.325 \\ 
     \textsc{NGram} & 0.056 & \textbf{0.132} & \textbf{4.892} & 0.502 & 5.582 & 0.126 & \textbf{0.235} & \textbf{3.196} & 0.178 & 3.735 \\  
        \bottomrule
    \end{tabular}
\end{table*}

\textsc{NGram} has lower $d_c$ and $d_t$ values than all other methods, although it performs less well (comparatively) when analyzing $d_s$.  
This indicates that, although the category and time dimensions of the STC region merging seem well-suited, the spatial merging may be too coarse.  
Less merging in the spatial dimension would help to minimize accuracy losses here, although a moderate decrease in efficiency would have to be tolerated.
Space limitations prohibit deeper analysis of different STC region merging approaches.

\subsubsection{Runtime Analysis}
\label{sss:runtime}

Table \ref{tab:runtime-analysis} shows the average runtime of each perturbation method, including a breakdown of time spent on each stage of the mechanism. 
The `Other' column incorporates overheads and mechanism stages unique to one perturbation method (e.g., time smoothing in \textsc{IndNoReach} and \textsc{IndReach}, or the POI-level reconstruction in \textsc{NGramH}).  
As expected, \textsc{IndNoReach} and \textsc{IndReach}  are exceptionally quick as they rely solely on indexing operations.  
For the remaining mechanisms, the majority of the runtime is reserved for solving the optimization problem during the trajectory reconstruction phase.
All other phases are performed in sub-second times.  
This demonstrates that even quicker results are feasible if time is spent selecting the best LP solver and tuning the optimization parameters -- aspects of work that were beyond the scope of this paper.
Importantly, however, \textsc{NGram} complements its accuracy superiority with efficiency prowess over \textsc{NGramNoH} and \textsc{PhysDist}, being nearly two and four times faster on average, respectively.  
The performance gain is primarily achieved from having a smaller optimization problem as a result of STC region merging.

\subsection{Parameter Variation}
\label{ss:param-variation}
We now examine how performance is influenced by the trajectory characteristics, and the mechanism or experiment parameters.
Figures \ref{fig:error-graphs} and \ref{fig:runtime-graphs} show the variation in NE and runtime, respectively.

\subsubsection{Trajectory Length}
\label{sss:trajLen}
Figures \ref{fig:trajlen-error-taxi} and \ref{fig:trajlen-error-SG} show an increase in error as trajectory length increases.  
\textsc{NGram} consistently outperforms other methods, which are broadly comparable in accuracy terms, with the exception of \textsc{PhysDist}.  
This is because $\epsilon'$ decreases as $|\tau|$ increases, which decreases the likelihood that the true $n$-gram is returned.  
Although the reconstruction stage seeks to minimize the effects of this, the reconstruction error is defined with respect to the \textit{perturbed} $n$-grams (not the \textit{real} $n$-grams), which limits the ability for the mechanism to correct itself.  
An alternative privacy budget distribution would be to assign a constant $\epsilon'$ value for each perturbation, but this means trajectories experience a different amount of privacy leakage (i.e., $\epsilon = (n+|\tau|-1)\epsilon'$).  

Figures \ref{fig:trajlen-runtime-taxi} and \ref{fig:trajlen-runtime-SG} show how the runtime changes with trajectory length.  
As expected, \textsc{IndNoReach} and \textsc{IndReach} show little runtime variability.  
Of the optimization-based approaches, \textsc{NGram} is consistently the fastest method, and its rate of increase as $|\tau|$ increases is lower than other approaches.  
Finally, as most trajectories were less than eight POIs in length, \textsc{NGram} produces output trajectories in a reasonable time for the vast majority of trajectories.

\subsubsection{Privacy Budget}
\label{sss:privacy-budget}
\begin{figure}[!t]
    \centering
    \begin{subfigure}[b]{\columnwidth}
    \centering
        \includegraphics[height = 0.4cm]{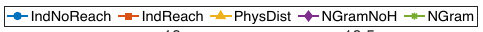}
        \label{fig:legend}
    \end{subfigure}
    \\
    \begin{subfigure}[b]{0.32\columnwidth}
        \centering
        \includegraphics[height = 3cm]{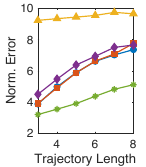}
        \caption{Traj. Length (T-F)}
        \label{fig:trajlen-error-taxi}
    \end{subfigure}
    \hfill
    \begin{subfigure}[b]{0.32\columnwidth}
        \centering
        \includegraphics[height = 3cm]{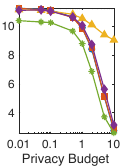}
        \caption{Priv. Budget (T-F)}
        \label{fig:pb-error-taxi}
    \end{subfigure}
    \hfill
    \begin{subfigure}[b]{0.32\columnwidth}
        \centering
        \includegraphics[height = 3cm]{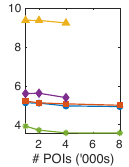}
        \caption{Size of $|\mathcal{P}|$ (T-F)}
        \label{fig:n-error-taxi}
    \end{subfigure}
    \\
    \begin{subfigure}[b]{0.32\columnwidth}
        \centering
        \includegraphics[height = 3cm]{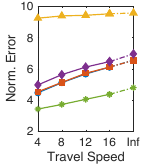}
        \caption{Travel Speed (T-F)}
        \label{fig:theta-error-taxi}
    \end{subfigure}
    \hfill
    \begin{subfigure}[b]{0.32\columnwidth}
        \centering
        \includegraphics[height = 3cm]{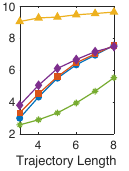}
        \caption{Traj. Length (SG)}
        \label{fig:trajlen-error-SG}
    \end{subfigure}
    \hfill
    \begin{subfigure}[b]{0.32\columnwidth}
        \centering
        \includegraphics[height = 3cm]{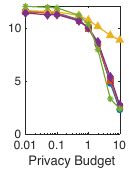}
        \caption{Priv. Budget (SG)}
        \label{fig:pb-error-SG}
    \end{subfigure}
    \\
    \begin{subfigure}[b]{0.32\columnwidth}
        \centering
        \includegraphics[height = 3cm]{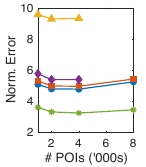}
        \caption{Size of $|\mathcal{P}|$ (SG)}
        \label{fig:n-error-SG}
    \end{subfigure}
    \hfill
    \begin{subfigure}[b]{0.32\columnwidth}
        \centering
        \includegraphics[height = 3cm]{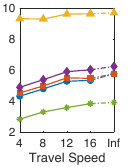}
        \caption{Travel Speed (SG)}
        \label{fig:theta-error-SG}
    \end{subfigure}
    \hfill
    \begin{subfigure}[b]{0.32\columnwidth}
        \centering
        \includegraphics[height = 3cm]{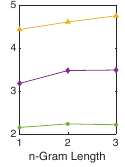}
        \caption{$n$-gram Length}
        \label{fig:ngram-error}
    \end{subfigure}
    \caption{Normalized error as experimental settings vary; Figures \ref{fig:trajlen-error-taxi}-\ref{fig:theta-error-taxi} use Taxi-Foursquare (T-F) data, Figures \ref{fig:trajlen-error-SG}-\ref{fig:theta-error-SG} use Safegraph (SG) data, and Figure \ref{fig:ngram-error} uses Campus data}
    \label{fig:error-graphs}
\end{figure}

Figures \ref{fig:pb-error-taxi} and \ref{fig:pb-error-SG} show how NE is influenced by $\epsilon$.  
All methods produce expected error profiles---as $\epsilon$ increases, the error decreases---although this behavior is less notable for \textsc{PhysDist}.  
When $\epsilon < 1$, the drop-off in utility is less pronounced.
This behavior is likely to be indicative of the DP noise overwhelming the characteristics of the true data (i.e., the output data offers little value due to the added noise).
Hence, we recommend setting $\epsilon \geq 1$ as a `starting point' in applications of our solution.

\begin{figure}[!t]
    \centering
    \begin{subfigure}[b]{\columnwidth}
    \centering
        \includegraphics[height = 0.4cm]{Figures/legend.pdf}
        \label{fig:legend}
    \end{subfigure}
    \\
    \begin{subfigure}[b]{0.32\columnwidth}
        \centering
        \includegraphics[height = 3cm]{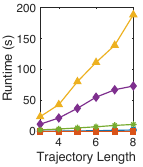}
        \caption{Traj. Length (T-F)}
        \label{fig:trajlen-runtime-taxi}
    \end{subfigure}
    \hfill
    \begin{subfigure}[b]{0.32\columnwidth}
        \centering
        \includegraphics[height = 3cm]{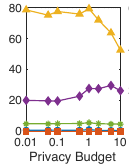}
        \caption{Priv. Budget (T-F)}
        \label{fig:pb-runtime-taxi}
    \end{subfigure}
    \hfill
    \begin{subfigure}[b]{0.32\columnwidth}
        \centering
        \includegraphics[height = 3cm]{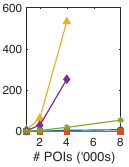}
        \caption{Size of $|\mathcal{P}|$ (T-F)}
        \label{fig:n-runtime-taxi}
    \end{subfigure}
    \\
    \begin{subfigure}[b]{0.32\columnwidth}
        \centering
        \includegraphics[height = 3cm]{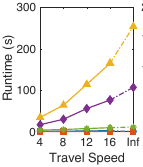}
        \caption{Travel Speed (T-F)}
        \label{fig:theta-runtime-taxi}
    \end{subfigure}
    \hfill
    \begin{subfigure}[b]{0.32\columnwidth}
        \centering
        \includegraphics[height = 3cm]{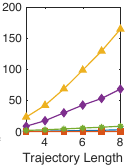}
        \caption{Traj. Length (SG)}
        \label{fig:trajlen-runtime-SG}
    \end{subfigure}
    \hfill
    \begin{subfigure}[b]{0.32\columnwidth}
        \centering
        \includegraphics[height = 3cm]{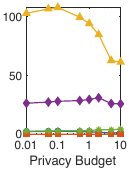}
        \caption{Priv. Budget (SG)}
        \label{fig:pb-runtime-SG}
    \end{subfigure}
    \\
    \begin{subfigure}[b]{0.32\columnwidth}
        \centering
        \includegraphics[height = 3cm]{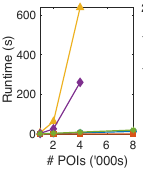}
        \caption{Size of $|\mathcal{P}|$ (SG)}
        \label{fig:n-runtime-SG}
    \end{subfigure}
    \hfill
    \begin{subfigure}[b]{0.32\columnwidth}
        \centering
        \includegraphics[height = 3cm]{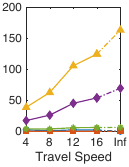}
        \caption{Travel Speed (SG)}
        \label{fig:theta-runtime-SG}
    \end{subfigure}
    \hfill
    \begin{subfigure}[b]{0.32\columnwidth}
        \centering
        \includegraphics[height = 3cm]{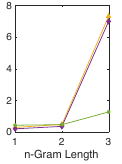}
        \caption{$n$-gram Length}
        \label{fig:ngram-runtime}
    \end{subfigure}
    \caption{Average runtime as experimental settings  vary; Figures \ref{fig:trajlen-runtime-taxi}-\ref{fig:theta-runtime-taxi} use Taxi-Foursquare (T-F) data, Figures \ref{fig:trajlen-runtime-SG}-\ref{fig:theta-runtime-SG} use Safegraph (SG) data, and Figure \ref{fig:ngram-runtime} uses Campus data}
    \label{fig:runtime-graphs}
\end{figure}

Importantly, Figures \ref{fig:pb-runtime-taxi} and \ref{fig:pb-runtime-SG} show that the runtime of \textsc{NGram} is relatively immune to the privacy budget, which indicates that the scale of the optimization problem is unaffected by the privacy budget.  
Most remaining methods also exhibit this immunity, although $\textsc{PhysDist}$ does not, which further emphasizes the benefits of including external information in trajectory perturbation.

\subsubsection{Size of POI Set}
\label{sss:n-pois}
Figures \ref{fig:n-error-taxi} and \ref{fig:n-error-SG} show the effect that $|\mathcal{P}|$ has on NE.  
We omit \textsc{PhysDist} and \textsc{NGramNoH} when $|\mathcal{P}| = 8{,}000$, owing to their high runtime.  
Interestingly, the error profiles are relatively immune to the effects of changing $|\mathcal{P}|$.  
This suggests that the optimal reconstruction phase can effectively identify the best trajectory from the perturbed $n$-grams.
Figure \ref{fig:n-runtime-taxi} and \ref{fig:n-runtime-SG} show a moderate runtime increase for \textsc{NGram}, which still perturbs trajectories in a reasonable time, even for large POI sets.  
In all $n$-gram-based methods, at least 95\% of runtime is spent during reconstruction, indicating the area of focus if substantial time-savings are necessary.

\subsubsection{Reachability Constraint}
\label{sss:reachability-results}
We experiment with assumed travel speeds of $\{4,8,12,16\}$km/hr, and we also consider imposing no reachability constraint (i.e., $\theta = \infty$).
Error increases as the reachability constraint becomes less strict or is removed entirely (Figures \ref{fig:theta-error-taxi} and \ref{fig:theta-error-SG}).
This is because more $n$-grams are feasible and so the likelihood of the true $n$-gram being returned is reduced.
\textsc{NGram} consistently outperforms all other methods in accuracy terms and, in terms of runtime, it is relatively immune to changes in assumed travel speed, unlike other $n$-gram approaches (Figures \ref{fig:theta-runtime-taxi} and \ref{fig:theta-runtime-SG}).
Importantly, \textsc{NGram} is up to 31\% better than other methods when the reachability constraint is applied, and it remains up to 22\% better than other methods when the reachability constraint is omitted.

\subsubsection{$n$-gram Length}
\label{sss:ngram-length}
We consider $n$-grams of length $\{1,2,3\}$ for the three $n$-gram-based methods, using the Campus data.  
The NE and runtime results are shown in Figures \ref{fig:ngram-error} and \ref{fig:ngram-runtime}, respectively. 
\textsc{NGram} consistently outperforms other methods for all values of $n$, and, for \textsc{NGram}, $n = 2$ offers the best results.  
This is to be expected given the trade-off between capturing more information between neighboring points (achieved with high $n$) and the division of $\epsilon$ and sensitivity of the distance function (where low $n$ is good).  
As expected and discussed in Section \ref{ss:efficiency-discussion}, runtime costs start to become undesirable when $n=3$, supporting our recommendation that bigrams should be used in most real-world applications.

\subsection{Application-Inspired Queries}
\label{ss:application-queries}
\begin{figure}[t]
\centering
    \begin{subfigure}[b]{\columnwidth}
    \centering
        \includegraphics[height = 0.4cm]{Figures/legend.pdf}
        \label{fig:legend}
    \end{subfigure}
    \\
    \begin{subfigure}[b]{0.32\columnwidth}
        \centering
        \includegraphics[height=3cm]{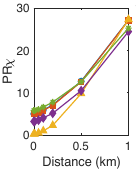}
        \caption{Space}
        \label{fig:range-space}
    \end{subfigure}
    \hfill
    \begin{subfigure}[b]{0.32\columnwidth}
        \centering
        \includegraphics[height=3cm]{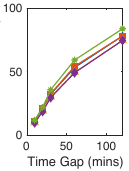}
        \caption{Time}
        \label{fig:range-time}
    \end{subfigure}
   \hfill
    \begin{subfigure}[b]{0.32\columnwidth}
        \centering
        \includegraphics[height=3cm]{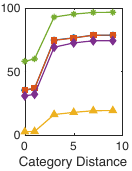}
        \caption{Category}
        \label{fig:range-cat}
    \end{subfigure}
    \caption{Variation in $\bm{PR_{\chi}}$ values as $\bm{\delta_\chi}$ changes}
    \label{fig:range}
\end{figure}

Figure \ref{fig:range} shows the results for the PRQs for each dimension.  
For space and time PRQs, all methods perform similarly, although \textsc{NGram} slightly outperforms the other methods in general.  
There is a more notable difference in performance for category PRQs, with \textsc{NGram} clearly superior for all $\delta_c$ values.  
Interestingly, there is an evident step at $\delta_c = 3.5$, which suggests strong preservation of category within levels 2 and 3, which demonstrates robustness in our solution.
The ability to preserve the general category of POIs indicates the solution's suitability for societal contact tracing as relevant agencies can, say, advise people who have recently visited sports stadia to monitor their health.
Table \ref{tab:hotspots} shows the AHD and ACD values under default settings (AHD is in hours, ACD has no units).  
\textsc{NGram} is much better than other methods in preserving the temporal location of hotspots. 
Again, \textsc{PhysDist} performs worst of all and the remaining methods are broadly comparable.  
Interestingly, \textsc{NGram} performs less well when considering ACD values.  
This suggests that, while hotspots are broadly preserved in time, they are `flatter' in the perturbed trajectory sets.  
In practice, preserving the spatio-temporal location of hotspots will probably be more important to policy makers and researchers than preserving the hotspot strength (i.e., the maximum number of unique visitors).

Based on these results and those in Section \ref{sss:privacy-budget}, we recommend $\epsilon \geq 1$ for most practical applications.

\begin{table}[tb]
    \centering
    \caption{AHD and ACD values for default trajectory sets}
    \label{tab:hotspots}
    \begin{tabular}{c|cc|cc|cc}
        \toprule
        \multirow{2}{*}{\textbf{Method}} & \multicolumn{2}{c|}{\textbf{Taxi}} & \multicolumn{2}{c|}{\textbf{Safegraph}} & \multicolumn{2}{c}{\textbf{Campus}} \\
        & \textbf{AHD} & \textbf{ACD} & \textbf{AHD} & \textbf{ACD} & \textbf{AHD} & \textbf{ACD} \\
        \midrule
            \textsc{IndNoReach} & 1.58 & 8.21 & 2.52 & 13.07 & 2.36 & \textbf{15.72} \\ 
            \textsc{IndReach} & 1.72 & 9.64 & 2.54 & \textbf{9.07} & 2.54 & 17.83 \\ 
            \textsc{PhysDist} & 2.22 & 10.76 & 3.34 & 16.24 & 4.38 & 23.48 \\ 
            \textsc{NGramNoH} & 1.71 & \textbf{9.36} & 2.81 & 11.25 & 3.29 & 18.23 \\ 
            \textsc{NGram} & \textbf{1.49} & 13.53 & \textbf{2.01} & 16.30 & \textbf{2.03} & 18.74 \\ 
        \bottomrule
    \end{tabular}
\end{table}

\section{Further Work}
\label{s:conc}
We have developed an efficient and scalable $n$-gram-based method for perturbing trajectory data in accordance with LDP.
However, there are a number of areas in which our solution can be extended, and these form the basis for future work for us and others.  
Although the external knowledge we use is limited to the data that is widely available, our framework can accommodate other data sources without difficulty.
For example, temporally-varying POI popularity and POI-specific opening hours can be incorporated easily.  Less-structured data (e.g., inferred popularity from public comments) could also be incorporated into the semantic distance function.
We anticipate that incorporating more, richer data sources would further enhance utility, without negatively affecting efficiency.  
POI attributes can also be extended to a personalized LDP setting (e.g., the privacy level of a hospital differs if one is a doctor or patient).

Whereas the focus of this paper has been on devising a general approach for trajectory sharing, our solution can be adapted for specific applications or to consider the setting where single location points are shared continuously.  
Applications with specific utility aims may necessitate tuning of parameters or distance functions which is another direction of future work. 
Our problem framework and solution can also be applied to any notion of trajectory in space-time.
To illustrate this, consider sharing shopping habits (e.g., credit card transactions).  
Here, $\mathcal{P}$ represents the set of purchasable products, with attributes such as category, price, etc.  
We can exploit intrinsic hierarchies such that $\mathcal{R}_c$ represents the set of stores from which products are purchased (which can be online or physical stores).  
The reachability constraint remains to ensure that adjacent stores in $\tau$ are reachable in the real world (as is currently done to identify and prevent credit card fraud).  Online stores would always be `reachable' given their non-physical presence.
Other concepts, such as utility-enhancing semantic distance functions and the impossibility of some combinations (e.g., purchasing a car from a florist), translate naturally. 
Hence, this framework can be applied more generally. 

\balance

\begin{acks}
This work is supported by the UK Engineering and Physical Sciences Research Council (EPSRC) under Grant No. EP/L016400/1, European Research Council under Grant No. ERC-2014-CoG 647557, and by The Alan Turing Institute under EPSRC Grant No. EP/N510129/1.  
The authors would also like to acknowledge the University of Warwick Scientific Computing Research Technology Platform for assistance in the research described in this paper.
\end{acks}

\pagebreak

\bibliographystyle{ACM-Reference-Format}
\bibliography{99-bib.bib}

\end{document}